# A novel method for monitoring proton beam spot using back-streaming secondary neutrons from the target

Xiaoyun Yang,[1,2,3] Hantao Jing,[2,3,*] Binbin Tian,[2,3,4,*] Li Ou,[1] Yankun Sun,[2,3] Xiaolong Gao,[2,3,5]

[1] *Guangxi Normal University, Guilin, 541004, China*
[2] *Institute of High Energy Physics, Chinese Academy of Sciences (CAS), Beijing 100049, China*
[3] *Spallation Neutron Source Science Center, Dongguan 523803, China*
[4] *School of Energy and Power Engineering, Xi'an Jiaotong University, Xi'an 710049 China*
[5] *Hebei Normal University, Shijiazhuang 10094, China*

[*] Corresponding authors
E-mail addresses: jinght@ihep.ac.cn (Jing Hantao) and tianbb@ihep.ac.cn (Tian Binbin)

**Abstract**
The power of the proton beam of a high-power spallation neutron source generally ranges from 100 kW to several MW. The distribution of the power density of the beam on the target is critical for the stable operation of the high-power spallation target. This study proposes a beam monitoring method that involves restoring the image of a high-power proton beam spot on a target based on the principle of pinhole imaging by using the back-streaming of secondary neutrons from the spallation target. Fast and indirect imaging of the beam spot can be achieved at a distance of tens of meters from the target. The proposed method of beam monitoring can flexibly adjust the size of the pinhole and the measurement distance to control the intensity of flux of the secondary neutrons according to the demands of the detection system, which is far from the high-radiation target area. The results of simulations showed that the proposed method can be used to restore the beam spot of the incident proton by using the point response function and images of the secondary neutrons. Based on the target and the Back-n beamline in the CSNS, the effectiveness of this method has also been confirmed.

## I. INTRODUCTION

Spallation neutron sources [1] and accelerator-driven subcritical systems (ADS) [2] are large facilities driven by high-energy and high-power proton beams. A large number of neutrons are released through spallation reactions brought about by high-energy and high-power proton beams bombarding heavy metal targets. For the spallation of neutron sources, the power of the proton beam generally varies from a few hundreds of kW to a few MW, and the ADS has an expected proton power of tens of MW. The distribution of the proton beam on the target is often uncertain, and varies qualitatively between a uniform and a Gaussian distribution. To match the shape of the spallation target with the beam and reduce the maximum power density at the center of the beam spot, the last proton beam transport system in front of the target usually is needed to handle such characteristics of the beam as spot homogenization, expansion in size, and changes in its distribution and shape. This changes the distribution of the proton beam, and further increases the uncertainty and the distribution of power density in front of the target [3]. However, the distribution and stability of the spot of proton beam on the target is critical to the stable operation of the high-power spallation target.

Therefore, the accurate monitoring and measurement of the two-dimensional (2D) distribution and power density distribution of the proton beam on the target is a common issue encountered in high-power spallation neutron sources and ADS systems worldwide.

A multifilament beam monitor installed in front of the spallation target is commonly used to measure the one-dimensional (1D) distribution of the proton beam in two directions. This method has been used at the Japan Spallation Neutron Source (JSNS) [4, 5], Spallation Neutron Source (SNS) [6], and China Spallation Neutron Source (CSNS) [7]. As the beam power is increased, the multifilament beam monitor faces an increased risk of exposure to the high-radiation environment. Another method that involves coating the surface of the target with luminescent material has been used by the SNS [8, 9], CSNS [10, 11], and European Spallation Source (ESS) [12]. The fluorescence emitted by the interaction between the proton beam and luminescent material is captured by the camera to achieve images of the beam spot. However, the luminescence efficiency will be greatly reduced, such that the luminescent material cannot be used for a long time. Moreover, for methods of imaging the beam based on fluorescence, the camera cannot be far from the target zone with very strong radiation. An attempt at the JSNS was made to place the activated material before the target for a while and subsequently use activation analysis to indirectly obtain the distribution of the protons once the accelerator had been shut down [4]. A clear disadvantage of this is that it cannot be used to monitor the system in real time. It is also difficult and time consuming to handle highly radioactive material.

This paper proposes a method to monitor the proton beam spot through back-streaming secondary neutron imaging. This involves obtaining the distribution of secondary neutrons at a distance through pinhole imaging and then restoring the distribution of the spot of proton beam through image reconstruction. This method overcomes the difficulties caused by radiation damage induced by a high-power proton beam, can be used to monitor the spot of the proton beam in real time, and is easy to maintain. It is expected to be an effective method to monitor high-power proton beam spots on the target.

## II. IDEA AND GEOMETRICAL STRUCTURE

To avoid radiation damage in a high-power accelerator–target complex owing to back-streaming secondary particles from the target and protect the components of the proton beamline, the latter is deflected by an angle of 15 degrees 20 meters from the target in the CSNS [13]. A back-streaming white neutron beamline (Back-n) [14] has been constructed in the backward direction for research in nuclear experimental physics. We now have the conditions needed to carry out the pinhole imaging of a spot of proton beam on the target by using back-streaming secondary neutrons. Construction is also underway based on a similar design philosophy at the ESS [15] and China initiative Accelerator Driven System (CiADS) [16] to deflect proton beamlines in front of a target.

The high-power proton beam is deflected at a certain angle before bombarding the spallation target. The interaction between the proton beam and the spallation target causes many complex nuclear reaction processes and produces a large number and variety of secondary particles, such as low-energy protons, neutrons, gamma rays, positrons, electrons, and pions. The intensity of outgoing particles through backward streaming can reflect the density distribution of the spot of the incident proton beam. Neutrons have the highest intensity of all these secondary particles. Furthermore, they are not affected by the magnetic focusing lattice for the proton beam. Therefore, secondary neutrons are first selected. The distribution of the back-streaming secondary neutrons is imaged at a distance by using pinhole imaging. An image reconstruction algorithm is used to restore the distribution of the spot of proton beam. The geometric scheme of the imaging of the beam spot using secondary neutrons is shown in Fig. 1. We focus here on the method of imaging in this case and research its feasibility. We assume that a high-power proton beam with an energy of 1.6 GeV bombards a cylindrical tungsten target. For the optics of pinhole imaging, the distance between the target surface and the pinhole is assumed to be equal to that between the detector and the pinhole.

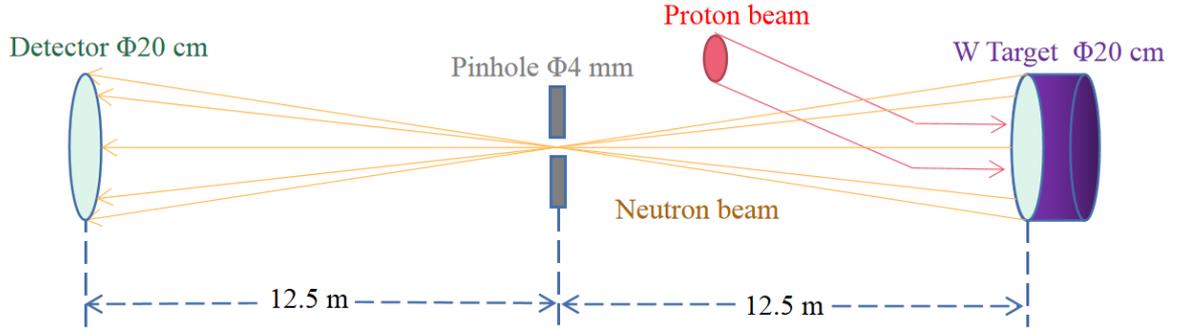

FIG. 1 Imaging schematics of spot of proton beam on a target.

### III. CHARACTERISTICS OF BACK-STREAMING SECONDARY NEUTRONS

Complicated nuclear reactions occur when a high-energy proton beam bombards a tungsten target [17]. A nuclear cascading process occurs. The high-energy protons directly interact with nucleons in the nucleus to generate a large number of high-energy neutrons. The angular distribution of the neutrons is usually anisotropic. The neutron energy is also closely related to the angle of emission. The excited heavy nuclei subsequently prompt a fission reaction, and a large number of fast neutrons (about 0.1–10 MeV) are produced during the evaporation process. These neutrons are isotropic, and the accelerator-based spallation neutron source uses mainly them. The energy spectrum of a large number of back-streaming neutrons generated on the target in Fig. 2 was simulated using FLUKA (FLUktuierende KAskade), a fully integrated particle physics Monte Carlo simulation package [18]. The range of neutron energy is wide, and spans over several orders of magnitude. If the energy of the incident proton is higher than this, the energy spectrum of the neutron widens as well. Thermal neutrons and epithermal neutrons with a large outgoing angle are mainly generated by the slowing of neutrons in the target. The yield of these neutrons is usually significantly closely related to the structure of the target. The peak energy spectrum occurs in the range of energy of fast neutrons, which form about 90.14% of the total.

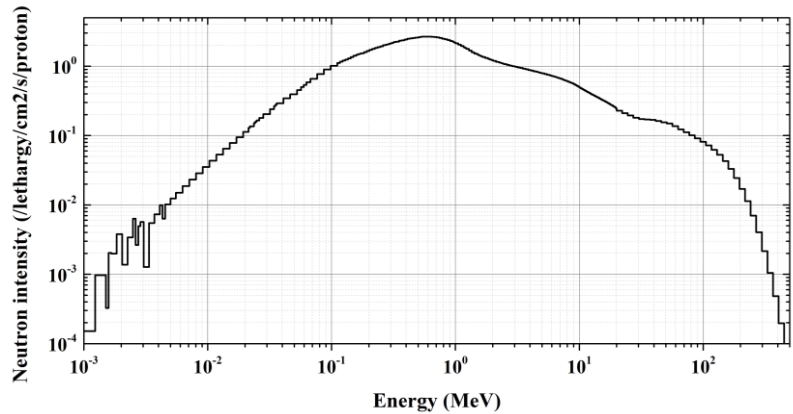

FIG. 2 Energy spectrum of neutron in the case of a 1.6 GeV proton beam bombarding a tungsten target.

Neutrons are highly penetrative. Secondary back-streaming neutrons backward-emitted from the target surface have a long range in the tungsten target. The distribution of the depth of neutrons in the target is shown in Fig. 3. The depth of most of neutrons is less than 10 cm from the target surface, and is affected by the structure of the target. The distributions of primary protons from a circular beam spot, with a diameter of Φ20 mm, and the distribution of the resulting secondary neutrons are shown in Fig. 4. The distribution of the neutron beam can adequately reflect the shape of the spot of the incident proton beam. Approximately 85% of the secondary neutrons are in the range of distribution of the spot of the primary proton beam.

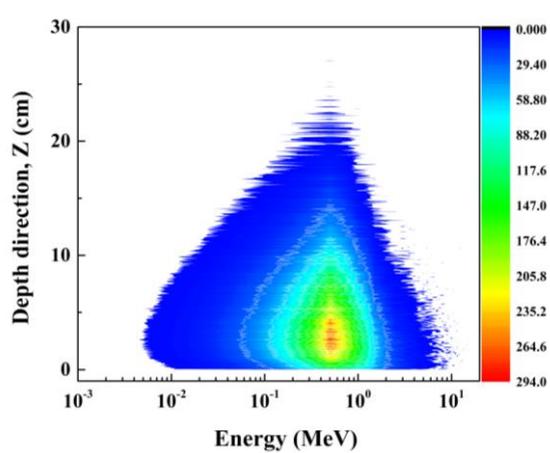

FIG. 3 Depth distribution of produced secondary neutrons with different energies in the target.

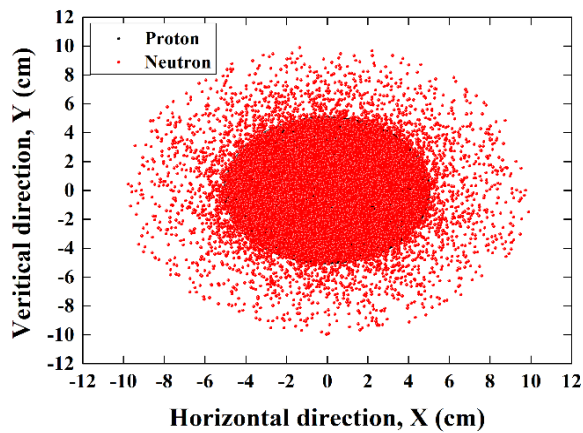

FIG. 4 Beam spots of primary protons and back-streaming secondary neutrons.

## IV. POINT RESPONSE FUNCTION

A high-energy proton produces about 40 neutrons on a tungsten target on average. As a whole, the distribution of back-streaming secondary neutrons on the front surface of the target is the superimposition of the probability distribution of neutrons produced by each proton hitting the target. The probability distribution of neutrons generated by a point source of protons is the key bond between the primary distribution of the proton beam and the distribution of the back-streaming neutron beam. A point response function (PRF) is defined as the probability distribution function of neutrons produced by a point source of protons. If the PRFs at different positions on the target are given, the distribution of the proton beam can be derived from the back-streaming neutrons.

For a spallation neutron source, the structures of the complex of high-power target, reflector, and moderator are complicated. In addition to neutron physics, such important factors as the cooling of the target and materials are also considered when a target station is designed. There are different structures of the spallation target along the longitudinal direction (direction of the proton beam) for different spallation targets. However, the transversal structure of the target is symmetrical about the direction of the proton beam. When designing of the target, the aiming area of the primary proton beam on the target surface is usually larger than the dimensions of the spot of the beam to ensure that it does not damage components outside the target. Therefore, we can assume that the probability distribution of neutrons produced by a point source of protons at different positions on the target surface is the same.

The probability distribution of back-streaming neutrons produced by every proton is determined by the complicated nuclear reactions and the longitudinal structure of the target. In this article, we simulated the 2D distribution of back-streaming neutrons emitted from a tungsten target, as shown in Fig. 5. The probability distribution function of the neutrons generated by a point source of protons was spike shaped and the range of dispersion at the bottom was small. The full-width at half-maximum (FWHM) of the profile was about 0.5 mm for the energy of larger than 600 keV. Compared with the usual spot of proton beam with dimensions of tens to hundreds of millimeters, the resolution of this point response distribution was sufficient to restore the primary distribution of the protons.

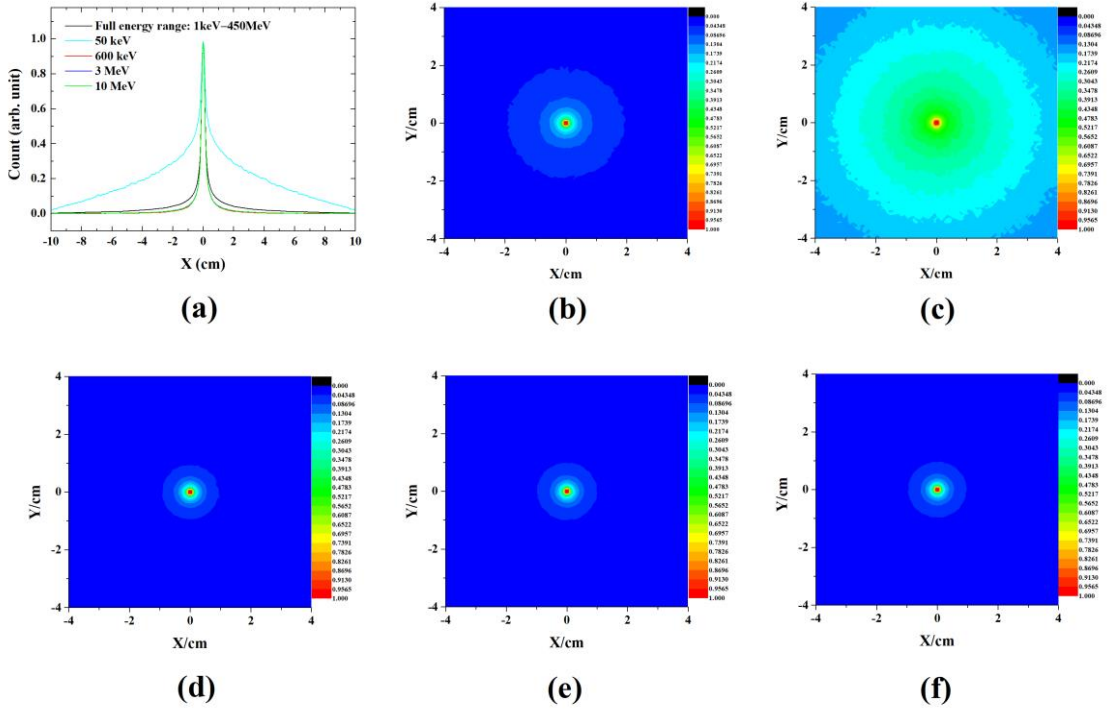

FIG. 5 Point response functions on the target surface for different ranges of neutron energy and the corresponding profile curves. (a) Profile curves of PRFs with different ranges of neutron energy, (b) full energy range, (c) 50 keV, (d) 600 keV, (e) 3 MeV, and (f) 10 MeV.

## V. RESTORATION OF PROTON BEAM SPOT
### A. Principle of image reconstruction

The production and transport of neutrons in the target involve complex nuclear reactions when high-energy protons bombard a heavy metal target. The distributions of the primary protons and the back-streaming neutrons from the target are inherently related. Assuming that the primary protons produce the same distribution of neutrons at each point they strike on the target surface, the distribution of the back-streaming neutrons is the convolution of the distribution of the primary proton beam and the PRF, namely:

$$N = PRF \otimes P, \qquad (1)$$

where $N$ represents the 2D distribution of back-streaming neutrons emitted from the target surface, $P$ represents the 2D distribution of primary protons, PRF is the point response function, and $\otimes$ represents the convolution operator. If a matrix form is used, the corresponding mathematical expression is as follows:

$$N = C \cdot P, \qquad (2)$$

where the matrix of the distribution of back-streaming neutrons can be expressed as $N = [n_{11}, n_{12}, ..., n_{ij}, ..., n_{IJ}]$, the distribution of the primary protons can be expressed as $P = [p_{11}, p_{12}, ..., p_{lm}, ..., n_{LM}]$, C is the response matrix formed by the point response function, and its matrix element is $C_{ij, lm}$. This represents the neutrons contributed by each proton point source in the matrix element $p_{lm}$ to the matrix element $n_{ij}$, and is written as:

$$C_{ij,lm} = \int_{\Delta x_{i,j}} \int_{\Delta y_{i,j}} PRF_{l,m}(x,y) dx dy, \qquad (3)$$

where $PRF_{l,m}(x, y)$ denotes the point response function located in the grid of $(l, m)$ as shown in Fig. 6. To neglect the influence of grid size and achieve better image resolution, the minimum bin sizes of the distributions of the neutron and proton were preferably smaller than or close to the FWHM of the point response function.

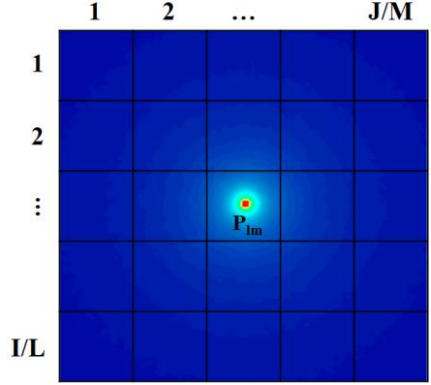

FIG. 6 Matrices of primary protons and back-streaming neutrons, and the point response function.

Based on Eq. (2), the correlation between the distribution of the back-streaming neutrons and the distribution of primary protons can be expressed by the corresponding linear equations:

$$\begin{pmatrix} n_{11} \\ \vdots \\ n_{ij} \\ \vdots \\ n_{IJ} \end{pmatrix} = \begin{pmatrix} C_{11,11} & \cdots & C_{11,lm} & \cdots & C_{11,LM} \\ \vdots & \ddots & \vdots & & \vdots \\ C_{ij,11} & \cdots & C_{ij,lm} & \cdots & C_{ij,LM} \\ \vdots & & \vdots & \ddots & \vdots \\ C_{IJ,11} & \cdots & C_{IJ,lm} & \cdots & C_{IJ,LM} \end{pmatrix} \begin{pmatrix} p_{11} \\ \vdots \\ p_{lm} \\ \vdots \\ p_{LM} \end{pmatrix} \quad (4)$$

In this paper, an iterative method, the Kaczmarz method [19], is used to solve the linear equations. It is specified by the following equation:

$$P_{l,m}(k+1) = P_{l,m}(k) + \lambda \frac{N_{i,j} - \sum_{l,m} C_{ij,lm} P_{lm}(k)}{\sum_{l,m} C_{ij,lm}^2} \quad (5)$$

To avoid the zone of the high-radiation target, we use the pinhole imaging method to obtain the distribution of back-streaming neutrons N' at a distance. According to the principle of pinhole imaging, there is a linear relation between the distribution of back-streaming neutrons N before the target and their remote distribution N', namely $N' = a \times N$, where $a$ is the attenuation coefficient due to pinhole collimation that is determined by the aperture and length of the imaging system.

**B. Image reconstruction**

In this section, the two typical functions of the uniform distribution and the Gaussian distribution are used to determine the distribution of the primary protons to verify the effectiveness of this research, respectively. A uniformly distributed proton beam over a square area of $10 \times 10$ cm$^2$ was first simulated to bombard the spallation target. The spots of back-streaming neutrons with different energies at full range, 50 keV, 600 keV, 3 MeV, and 10 MeV, were scored at a distance of 25 meters from the target by the pinhole imaging system, as shown in Figs. 7(a)–(e). The images of the restored spot of proton beam reconstructed from the distributions of the back-streaming neutrons (Figs. 7(a)–(e)) are shown in Figs. 7(f)–(j), respectively. Owing to the large divergence of neutrons with an energy of 50 keV (Figs. 5(a) and (c)) and interference from the surrounding low-energy background neutrons, we could not adequately reconstruct the distribution of the spot of primary proton beam, as shown in Fig. 7(g). The PRFs of the high-energy neutrons were much narrower. Therefore, the high-energy neutrons could adequately restore the distribution of the primary proton beam as shown in Figs. 7(c)–(e).

The same scenario for neutrons with the full energy range was also implemented as shown in Fig. 7(f), although the image of the restored proton beam spot in Fig. 7(a) is somewhat blurred. The fraction of neutrons with energy higher than 600 keV in the spectrum of the back-streaming neutrons was more than 55%. The contrast between the images of these restored proton spots shows that choosing higher-energy neutrons can yield an image of better quality.

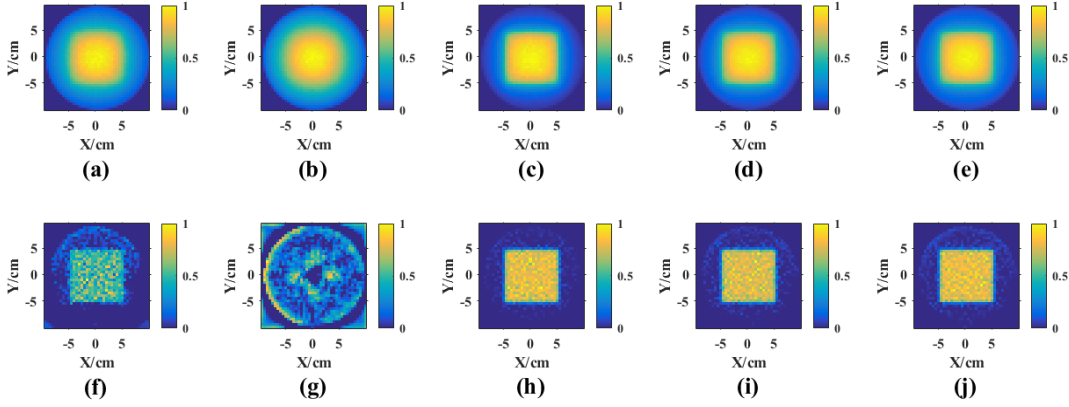

FIG. 7 Verification of the uniformly distributed proton beam. (a)–(e) Images of the distributions of the back-streaming neutrons 25 m from the target with different energies of full range, 50 keV, 600 keV, 3 MeV, and 10 MeV, respectively; (f)–(j) images of restored proton beam spot reconstructed from the distributions of the back-streaming neutrons corresponding to (a)–(e), respectively.

To verify the Gaussian distribution, a proton beam with a 2D Gaussian distribution (FWHMx = FWHMy = 6 cm) was used, as shown in Fig. 8(a). Neutrons with a peak energy of 600 keV were used to restore the spot of proton beam. The primary distribution of the Gaussian proton beam, the distribution of the back-streaming neutrons, and that of the restored proton beam are shown in Fig. 8. The horizontally projected curves are shown in Fig. 9. The FWHM difference between the primary proton beam and the restored proton beam was about 1.3%.

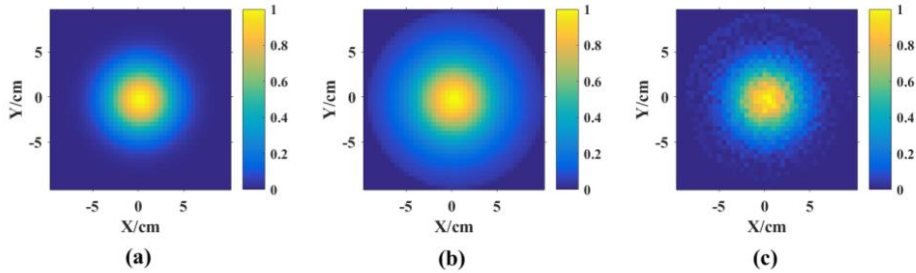

FIG. 8 (a) Verification of the Gaussian proton beam. (a) Distribution of the incident proton beam spot, (b) distribution of the back-streaming neutrons 25 m from the target at an energy of 600 keV, and (c) distribution of restored proton beam spot.

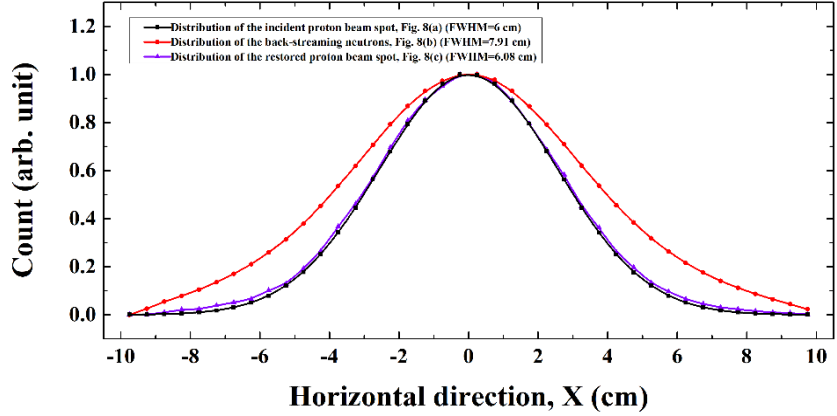

FIG. 9 Projected curves as a function of the horizontal coordinates corresponding to Figs. 8(a), (b), and (c) respectively.

### C. Simulation of CSNS proton beam spot on target

#### 1. Layout of CSNS before target

Figure 10 shows the layout of the CSNS target station, the primary proton beamline, and the back-streaming white neutron beamline. The proton beam was accelerated to 1.6 GeV by a multistage accelerator facility. Then, the 1.6 GeV proton beam was deflected by the last dipole and bombarded a heavy metal tungsten target. The spallation neutron sources used the spallation reaction between the high-energy protons and the target to generate a large number of neutrons for multidisciplinary research [20]. The complex of the target–moderator–reflector of CSNS is complicated. The target was composed of 11 layers of tantalum-coated tungsten sheets. The cross-sectional size of each tungsten sheet was $170 \times 70$ mm$^2$, and they were cooled by using water. The back-streaming white neutron source used a wide-energy neutron beam in the backward direction to carry out related research [21]. The Back-n had two experimental endstations, ES#1 and ES#2. A shutter is installed 31.27 m from the target surface on the neutron beamline that also functioned as a collimator. To satisfy the requirements of different experiments, several apertures, Φ3 mm, Φ 12 mm, Φ 50 mm, and 78 mm×62 mm could have been selected [22, 23]. Based on the condition of the pinhole on the Back-n neutron beamline, images of the proton beam spot before the target can be carried out using pinhole imaging.

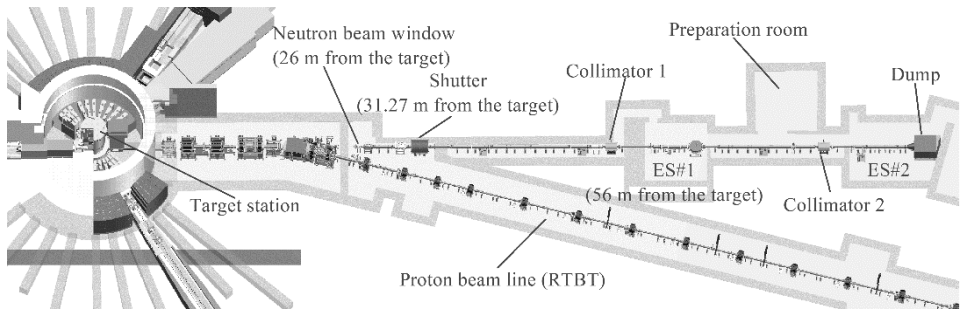

FIG. 10 Layout of CSNS target station, proton beamline, and Back-n beamline.

#### 2. Imaging of CSNS proton beam spot

FLUKA was used to simulate the nuclear reaction between the 1.6 GeV pencil proton beam and the tungsten target of the CSNS. Neutron distribution in the backward direction, namely, the point response function, was obtained. As shown in Fig. 11, the resolution ($\Delta E_{FWHM}/E_{peak}$) of the point response function was about 2.75 mm. Because the beam area to aim on the target was much larger than the proton beam spot, and the structure of the target at different positions on its surface along the direction of the proton beam was consistent, it was assumed that the point response function at any position on the target surface was the same. The CSNS proton beam spot matched the shape of the aiming area of the beam on the target, and had a rectangular distribution. The rectangular distribution of the protons is given in Fig. 12 by fitting the measured data from the multifilament beam monitor in front of the target [24].

The Back-n white neutron beamline and the Φ3 mm aperture on the shutter formed a pinhole imaging condition. We used the distribution of the back-streaming neutrons in ES#1 to restore the distribution of the primary protons. The distribution of the back-streaming neutrons on the target surface, neutron distribution in ES#1 through the small hole, and the distribution of the restored protons are given in Fig. 13. A comparison of projection curves between the restored distribution of the protons and the measured experimental data of the multifilament beam monitor before the target yielded good agreement, as shown in Fig. 14. But we also identified some background noise in the restored distribution of the protons. This was caused by multiple scatterings of neutrons in the target, and their effects of the transport beam tube, the wall of the aperture of the collimator, and the image reconstruction algorithm.

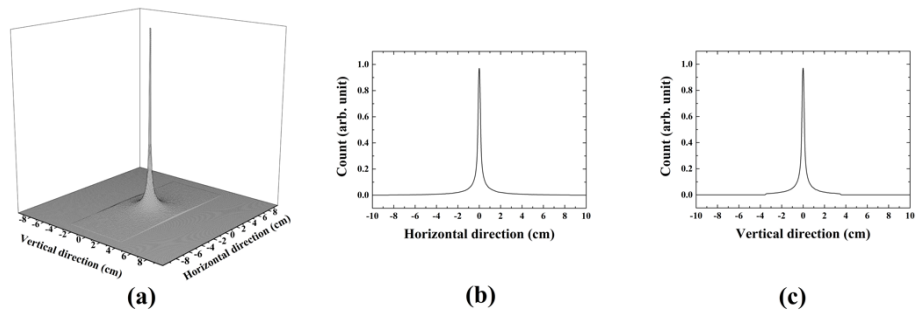

FIG. 11 Point response function corresponding to the CSNS-I target station. (a) 3D point response function, (b) horizontal profile, and (c) vertical profile.

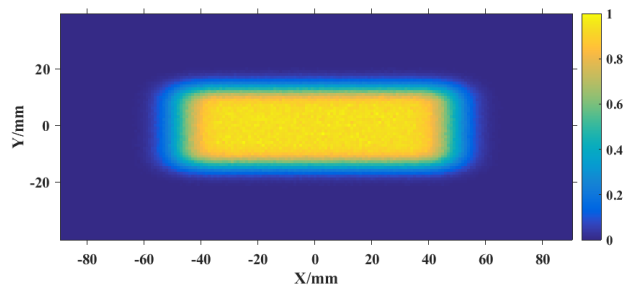

FIG. 12 Distribution of incident proton beam of CSNS before the target.

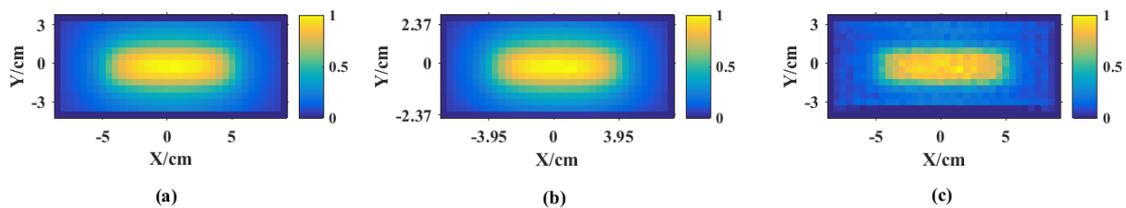

FIG. 13 Reconstructed results obtained by using a fitting distribution of the primary protons and back-streaming neutrons at an energy of 600 keV. (a) Distribution of the back-streaming neutrons before the target. (b) Neutron distribution after the pinhole system 56 m from the target. (c) Restored distribution of the spot of primary proton beam.

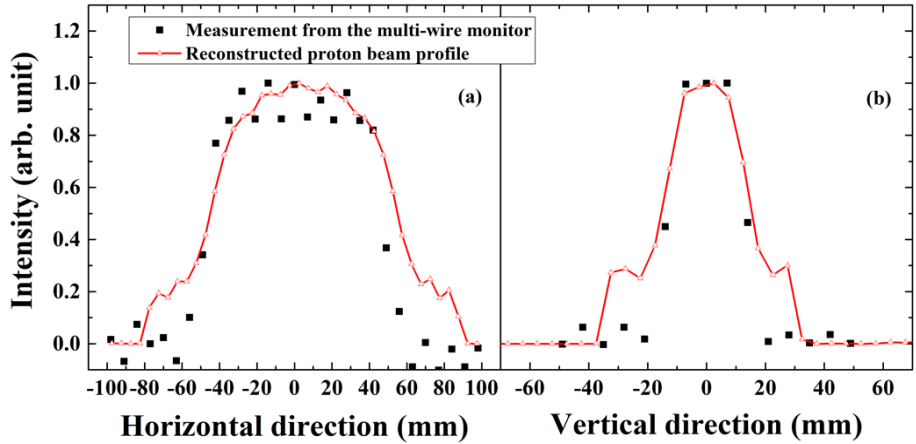

FIG. 14 Projected curves of the spot of the restored proton beam in two directions (red lines), and measured data from the multifilament monitor before the target (black dots).

## VI. CONCLUSIONS

This study proposed a method to image the spot of proton beam on a spallation target. By introducing a point response function, the relationship between the distribution of incident protons and that of back-streaming secondary neutrons was identified. The pinhole imaging method was used to measure the distribution of the back-streaming neutrons at a certain distance from the target of irradiation. The spot of proton beam on the target was then restored through image reconstruction. For spallation neutron sources worldwide and similar high-power neutron facilities, neutrons with energies above 600 keV in the backward direction can be used to restore the spot of primary proton beam. Higher-energy neutrons can deliver a better reconstruction.

The distribution of the proton beam can be completely restored by applying this method to the reconstruction of the spot of proton beam on the target in the CSNS. The restored proton distribution contains a small amount of background noise caused by scattering effects and the reconstruction algorithm. The proposed method of imaging the proton beam spot is thus feasible. The measurement avoids the area of strong radiation of the spallation target, especially for MW-level high-power spallation sources. In future work, we will carry out further experimental research to verify the Back-n.

**Acknowledgments**

This work was supported by the National Natural Science Foundation of China (Project: 12075135) and China postdoctoral science foundation (Project: 2021M691859).